\documentclass[11pt,a4p]{article}  

\usepackage{epsfig}

\def\etal{{\it et al.}}

\def\tm{\tau^-}
\def\tp{\tau^+}
\def\nut{\nu_{\tau}}

\def\gt{g_{\tau}}
\def\gm{g_{\mu}}
\def\ge{g_{\mathrm{e}}}

\def\etot{E_{\mathrm{tot}}}
\def\ptot{p_{\mathrm{tot}}}
\def\ecm{E_{\mathrm{cm}}}

\def\taue  {\tm \rightarrow \mbox{e}^- \bar{\nu}_{e}\nut }
\def\taum  {\tm \rightarrow \mu^- \bar{\nu}_{\mu} \nut }

\def\tauhpmay{\tau^- \rightarrow {\mbox {\rm h}}^- \geq 0\pi^0 \nu_{\tau}} 
\def\tauhhh{\tau^- \rightarrow \mathrm{h}^-\mathrm{h}^-\mathrm{h}^+ \geq 0\pi^0 \nu_{\tau}}

\def\ee{\mbox{e}^+\mbox{e}^-}
\def\mm{\mu^+\mu^-}
\def\tt{\tau^+\tau^-}
\def\qq{\mbox{q}\overline{\mbox{q}}}
\def\ggee{(\ee)\; \ee}
\def\ggmm{(\ee)\; \mm}
\def\ggtt{(\ee)\; \tt}

\def\reeee{\ee \rightarrow \ee}
\def\reemm{\ee \rightarrow \mm}
\def\reeqq{\ee \rightarrow \qq}

\def\reeeeee{\ee \rightarrow \ggee}
\def\reeeemm{\ee \rightarrow \ggmm}
\def\reeeett{\ee \rightarrow \ggtt}
\def\reeeeff{\ee \rightarrow (\ee)\; \mathrm{f\bar{f}}}

\def\dedxt{\mathrm{d}E/\mathrm{d}x}

\def\pa{p_{\mathrm{1}}}
\def\pb{p_{\mathrm{2}}}
\def\pc{p_{\mathrm{3}}}
\def\pmu{p_{\mathrm{\mu}}}
\def\popp{p_{\mathrm{1-opp}}}

\def\nmu{N_{\mathrm{muon}}}

\def\chisq{\chi^2}

\def\acol{\theta_{\mathrm{acol}}}
\def\mumatch{\mu_{\mathrm{match}}}
\def\ejet{E_{\mathrm{jet}}}

\def\btaum{B($\taum$)}
\begin{document}
\begin{titlepage}
\begin{center}{\large   EUROPEAN ORGANIZATION FOR NUCLEAR RESEARCH
}\end{center}\bigskip
\begin{flushright}
       CERN-EP/2002-085   \\ 15 November 2002
\end{flushright}
\bigskip\bigskip\bigskip\bigskip\bigskip
\begin{center}{\huge\bf   A Measurement of the \\ \boldmath $\taum$
    \unboldmath Branching Ratio
}\end{center}\bigskip\bigskip
\begin{center}{\LARGE The OPAL Collaboration
}\end{center}\bigskip\bigskip
\bigskip\begin{center}{\large  Abstract}\end{center}
The $\taum$ branching ratio has been measured using data collected
from 1990 to 1995 by the OPAL detector at the LEP collider.  The
resulting value of 
\[
\mbox{\btaum} = 0.1734 \pm 0.0009(stat) \pm 0.0006(syst)
\]
 has been used in conjunction with other OPAL
measurements to test lepton universality, yielding the coupling
constant ratios $\gm/\ge = 1.0005 \pm 0.0044$ and $\gt/\ge = 1.0031
\pm 0.0048$, in good agreement with the Standard Model prediction of
unity.  A value for the Michel parameter $\eta = 0.004 \pm 0.037$ has
also been determined and used to find a limit for the mass of the
charged Higgs boson, $m_{\mathrm{H}^{\pm}} > 1.28 \tan\beta$,  in the Minimal
Supersymmetric Standard Model.

\bigskip\bigskip\bigskip\bigskip
\bigskip\bigskip
\begin{center}{\large
(To be submitted to Physics Letters B)
}\end{center}
\end{titlepage}
\begin{center}{\Large        The OPAL Collaboration
}\end{center}\bigskip
\begin{center}{
G.\thinspace Abbiendi$^{  2}$,
C.\thinspace Ainsley$^{  5}$,
P.F.\thinspace {\AA}kesson$^{  3}$,
G.\thinspace Alexander$^{ 22}$,
J.\thinspace Allison$^{ 16}$,
P.\thinspace Amaral$^{  9}$, 
G.\thinspace Anagnostou$^{  1}$,
K.J.\thinspace Anderson$^{  9}$,
S.\thinspace Arcelli$^{  2}$,
S.\thinspace Asai$^{ 23}$
% ...
}\end{center}\bigskip
\bigskip
$^{  1}$School of Physics and Astronomy, University of Birmingham,
Birmingham B15 2TT, UK
\newline
$^{  2}$Dipartimento di Fisica dell' Universit\`a di Bologna and INFN,
I-40126 Bologna, Italy
\newline
$^{  3}$Physikalisches Institut, Universit\"at Bonn,
D-53115 Bonn, Germany
\newline
% ...
$^{  5}$Cavendish Laboratory, Cambridge CB3 0HE, UK
\newline
% ...
$^{  9}$Enrico Fermi Institute and Department of Physics,
University of Chicago, Chicago IL 60637, USA
\newline
% ...
$^{ 16}$Department of Physics, Schuster Laboratory, The University,
Manchester M13 9PL, UK
\newline
% ...
$^{ 22}$Department of Physics and Astronomy, Tel Aviv University,
Tel Aviv 69978, Israel
\newline
$^{ 23}$International Centre for Elementary Particle Physics and
Department of Physics, University of Tokyo, Tokyo 113-0033, and
Kobe University, Kobe 657-8501, Japan
\newline
% ...

\newpage
\section{Introduction}
Precise measurements of the leptonic decays of $\tau$ leptons provide
a means of stringently testing various aspects of the Standard Model.  OPAL previously has studied the leptonic $\tau$ decay modes by measuring the branching ratios  \cite{steve,opalt2m}, the Michel parameters \cite{rainer}, and radiative decays \cite{radt2m}.  This work presents a new OPAL measurement of the
$\taum$
branching ratio\footnote{Charge conjugation is assumed throughout this
paper.}, using $\ee$ data taken from 1990 to 1995 at energies near the Z$^0$
peak, corresponding to an integrated luminosity of approximately 170
pb$^{-1}$.  A pure sample of $\tt$ pairs is selected from the data
set, and then the fraction of $\tau$
jets in which the $\tau$ has decayed to a muon is determined.  This fraction is then
corrected for backgrounds and inefficiencies.  The  selection of $\taum$ candidates relies on only a few variables, each of which provides a highly
effective means of separating background events from signal events
while minimising 
systematic uncertainty.  This new measurement supersedes the previous
OPAL measurement of B$(\taum) = 0.1736 \pm 0.0027$ which was obtained
using data collected in 1991 and 1992, corresponding to an integrated
luminosity of approximately 39 pb$^{-1}$ \cite{opalt2m}.  

        OPAL \cite{opal} is a general purpose detector covering almost
        the
        full solid angle with approximate cylindrical symmetry about
        the $\ee$ beam axis\footnote{In the OPAL coordinate system, the e$^-$ beam
        direction defines the $+z$ axis, and the $+x$ axis points from the detector towards the centre of the LEP
        ring.  The polar angle $\theta$ is
        measured from the $+z$ axis and the azimuthal angle $\phi$ is
        measured from the $+x$ axis.}.  The following subdetectors are of
        particular interest in this analysis:  the tracking system,
        the electromagnetic calorimeter, the hadronic calorimeter, and
        the muon chambers.  The tracking system includes two vertex
        detectors, $z$-chambers,  and a
        large volume cylindrical tracking drift chamber
        surrounded by a solenoidal magnet which provides a magnetic
        field of 0.435 T.  This system is used
        to determine 
        the particle momentum and rate of energy
        loss.  The electromagnetic calorimeter consists of
        lead-glass blocks backed by photomultiplier tubes or
        photo-triodes for the detection of \u{C}erenkov radiation
        emitted by relativistic particles.  The
        instrumented magnet return yoke serves as a hadronic
        calorimeter, consisting of up to
        nine layers of limited streamer tubes sandwiching eight layers
        of iron, with inductive readout of the tubes onto large pads
        and aluminium strips.  In the central region of the
        detector, the calorimeters are 
        surrounded by four layers 
        of drift chambers for the detection of muons emerging from the
        hadronic calorimeter.  In each of the forward regions, the muon detector
        consists of four layers of limited streamer tubes arranged
        into quadrants which are transverse to the beam direction,
        and two ``patch'' sections which provide coverage in areas
        otherwise left without detector capabilities due to the
        presence of cables and support structures. 

Selection efficiencies and kinematic variable distributions for the
present analysis were
modelled using Monte Carlo simulated $\tp\tm$ event samples generated
with the KORALZ 4.02 package \cite{koralz} and the TAUOLA 2.0 library \cite{tauola}.  These events
were then passed through a full simulation of the OPAL detector \cite{opalsim}.
Background contributions from non-$\tau$ sources were evaluated using
Monte Carlo samples based on the following generators:  multihadron
events $(\reeqq)$ were simulated using JETSET 7.3 and JETSET 7.4 \cite{jetset},
$\reemm$ events using KORALZ \cite{koralz}, Bhabha events using
BHWIDE \cite{bhwide}, and two-photon events using VERMASEREN \cite{vermaseren}.
\section{\label{tausel} The \boldmath $\tp\tm$ \unboldmath selection}
        At LEP1, electrons and positrons were made to collide at
        centre-of-mass energies close to the Z$^0$ peak, producing
        Z$^0$ bosons at rest which 
        subsequently decayed into back-to-back pairs of leptons
        or quarks from which the $\tp\tm$ pairs were selected
        for this analysis.  These highly relativistic  $\tau$
        particles  decay in
        flight close to the interaction point, resulting in two highly-collimated, back-to-back jets
        in the detector. 

This analysis uses the  standard OPAL
        $\tt$  selection \cite{tausel}, with slight modifications
        to reduce Bhabha background in
        the $\tt$ sample \cite{lineshape}.  
The $\tp\tm$ selection requires that an event have two
        jets as defined by the cone algorithm in reference
        \cite{jetalgo}, with a cone half-angle of $35^{\circ}$.
        The average $|\cos\theta|$ of the two jets
        was required to be less than 0.91, in order to restrict the
        analysis to regions of the detector that are well understood. In addition,  fiducial cuts were applied to restrict the events to regions of the
detector with reliable particle information and with high particle
identification efficiency.  If a jet was determined to be within a region of
the detector associated with gaps between hadronic calorimeter
sectors, or dead regions in the muon chambers due to the support
structures of the detector, the entire event was removed from the
        $\tt$ sample.  In regions near the anode wire planes in the
        tracking chamber, high momentum particles may have their momentum
        incorrectly reconstructed, an effect that is not
        well-modelled by the Monte Carlo simulations.  Therefore,
        events containing particles which traverse the detector near
        the anode planes were also removed from the sample.

  The main sources of background to the $\tp\tm$ selection are
        Bhabha events, dimuon events, multihadron events, and
        two-photon events.  
For each type of background remaining in the $\tt$ sample, a variable
        was chosen in which the signal and background can be visibly
        distinguished.  The relative proportion of background was
        enhanced by loosening criteria which would normally remove
        that particular type of background from the sample, and/or by
        applying criteria to reduce the contribution from signal and
        to remove other types of background.  A comparison of the data and Monte Carlo
        distribution in a background-rich region was then used to assess the modelling of the
        background and to estimate the corresponding systematic error on the
        branching ratio.  The Monte Carlo simulation provides the overall
        shape of the background distribution, while the normalization
        is measured from the data.  In most cases, the Monte Carlo simulation was
        found to be consistent with the data.  When the data and Monte
        Carlo distributions did not agree, the Monte Carlo
        simulation  was adjusted to fit the data.  Uncertainties of
        4\% to 20\% in the background estimates were obtained from the
        statistical uncertainty in the normalization, including the
        Monte Carlo statistical error.  The following paragraphs discuss the measurement
        of each type of background in the $\tt$ sample.  

        Bhabha events, $\reeee$,  have two-particle final
        states and thus can mimic $\tt$ events.  They are
        characterized by two high-momentum tracks and large energy deposition in the electromagnetic
        calorimeter. The criteria used to reject the Bhabha background
        are identical to those used in the Z$^0$ lineshape analysis to
        reject Bhabha events in the $\tt$ sample \cite{lineshape},
        rather than the standard OPAL $\tt$ selection.
        The requirement $\etot + \ptot < 1.4 \ecm$,
        for $\tt$ pairs with an average $|\cos\theta|$ of less than
        0.7, was also 
        added in this analysis to further reduce the Bhabha background, where $\etot$
        is the sum of the energies of all the electromagnetic
        calorimeter clusters, $\ptot$ is the scalar sum of the
        momenta of all tracks, and $\ecm$ is the centre-of-mass
        energy.  The Bhabha background remaining in the
        $\tp\tm$ sample was measured by comparing the distributions of
        total scalar momentum and of total energy deposition between the data and the Monte Carlo
        simulation, where the Bhabha background has been enhanced by
        relaxing the criteria on $\etot$ and $\ptot$.  The fraction of residual Bhabha background in
        the $\tt$ sample was estimated to be $0.00305 \pm
        0.00027$.

        Dimuon events, $\reemm$, also have two particle final states
        with high momentum tracks, but little energy deposition in the
        electromagnetic calorimeter.  Dimuon events are removed from
        the $\tt$ sample by requiring $\etot + \ptot < 0.6 \ecm$ in
        cases where both jets exhibit muon characteristics.  The dimuon background remaining after the $\tt$ selection was determined by measuring
        the dimuon contribution to the scalar momentum distribution
        in the data
        and in the Monte Carlo simulation, where the dimuon background
        has been enhanced by relaxing the criterion on $\etot + \ptot$.  The fractional background in the $\tp\tm$ sample was estimated to be $0.00108 \pm
        0.00022$.

        At LEP1 energies, multihadron events, $\reeqq$, typically have
        considerably higher track and cluster multiplicities than
        $\tt$ events, and are removed from the $\tt$ sample by requiring low multiplicities.   In addition, the $\tau$ jets 
        are typically much more collimated than multihadron
        jets.  The distribution of the maximum angle between any
        good  track in the jet (see reference \cite{tausel} for the
        definition of a good track) and the jet direction was used to
        evaluate the agreement between the data
        and the Monte Carlo modelling of these events, where the
        multihadron background has been enhanced by modifying the
        multiplicity criteria.  This resulted in a
        fractional background estimate of $0.00377 \pm
        0.00015$.

        In two-photon events, $\reeeeff$, the final
        state $\ee$ pair has a small
        scattering angle and disappears down the beam pipe, leaving a
        pair of low energy fermions, usually $\mm$ or $\ee$, in the
        detector \footnote{Two-photon events with $\tau$ particles,
        $\reeeett$, are considered to be signal.}.  Since
        these particles do not result from the decay of the Z$^0$,
        they are not constrained to be emitted back-to-back.  The
        $\tt$ selection rejects them based upon their low energy and
        relatively high acollinearity, $\acol$\footnote{Acollinearity is the supplement of the
        angle between the two jets.}.  The acollinearity criterion was relaxed in
        order to enhance the two-photon background so that it could be
        measured.  Additionally, for $\reeeeee$ events, each jet was
        required to exhibit electron characteristics, while for
        $\reeeemm$ events, each jet was required to exhibit muon
        characteristics.  The acollinearity
        distribution in the data then was compared with that in the Monte
        Carlo simulation to evaluate the 
        backgrounds in the $\tt$ sample for $\reeeemm$ and $\reeeeee$
        events, corresponding to fractional background estimates of $0.00108 \pm 0.00054$ and $0.00157 \pm 0.00028$, respectively.

        The $\tt$ selection leaves a sample of 96,898 candidate $\tt$
        events, with a predicted fractional background of $0.01055
        \pm 0.00072$.  The backgrounds in the $\tt$ sample are
        summarized in Table \ref{ttausel}.  
\begin{table} 
\begin{center}
\begin{tabular} {|l|c|} \hline
Background      & Contamination \\ \hline
$\reeee$        & $0.00305 \pm 0.00027$  \\
$\reemm$        & $0.00108 \pm 0.00022$ \\
$\reeqq$        & $0.00377 \pm 0.00015$ \\
$\reeeemm$      & $0.00108 \pm 0.00054$ \\ 
$\reeeeee$      & $0.00157 \pm 0.00028$  \\ \hline \hline
Total           & $0.01055 \pm 0.00072$ \\ \hline
 \hline 
\end{tabular}
\caption{\label{ttausel}
Fractional backgrounds in the $\tt$ sample together with their estimated uncertainties.}
\end{center}
\end{table}

\section{\label{taumsel}The \boldmath $\taum$ \unboldmath selection}
After the $\tp\tm$ selection, each of the 193,796 candidate $\tau$ jets is analysed
individually to see
whether it exhibits the characteristics of the required $\taum$ signature. A muon from a $\tau$ decay will result in a track in the central
tracking chamber, little energy in the electromagnetic and hadronic
calorimeters, and a track in the muon chambers.  The $\taum$ selection
is based on information from the central tracking chamber
and the muon chambers.  Calorimeter information is not used in the
main selection, but instead is used to
create an independent $\taum$ control sample that is used to estimate
the systematic error in the selection efficiency.  
The branching ratio of the $\taum$ decay is inclusive of radiation in
the initial or final state \cite{pdg}, and so the $\taum$ selection retains decays that are accompanied
by a radiative photon or a radiative photon that has converted in the
detector into an $\ee$ pair.

The $\taum$ candidates are selected from jets with one to three tracks
in the tracking chamber, where the tracks are ordered according to
decreasing particle momentum.  The highest momentum track is
assumed to be the muon candidate.

\begin{figure}
\begin{center}
\mbox{\epsfig{file=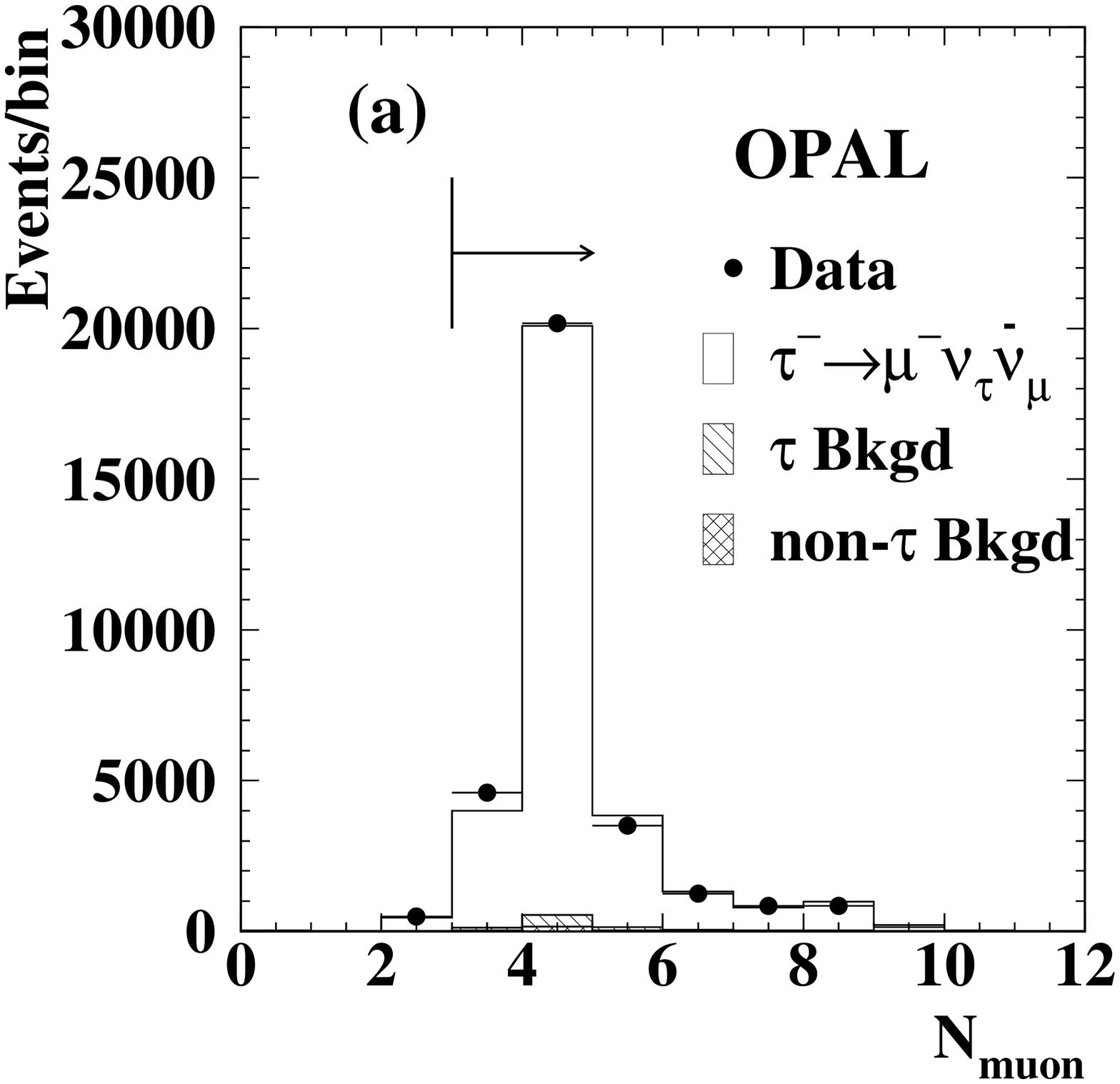,height=8cm}}
\mbox{\epsfig{file=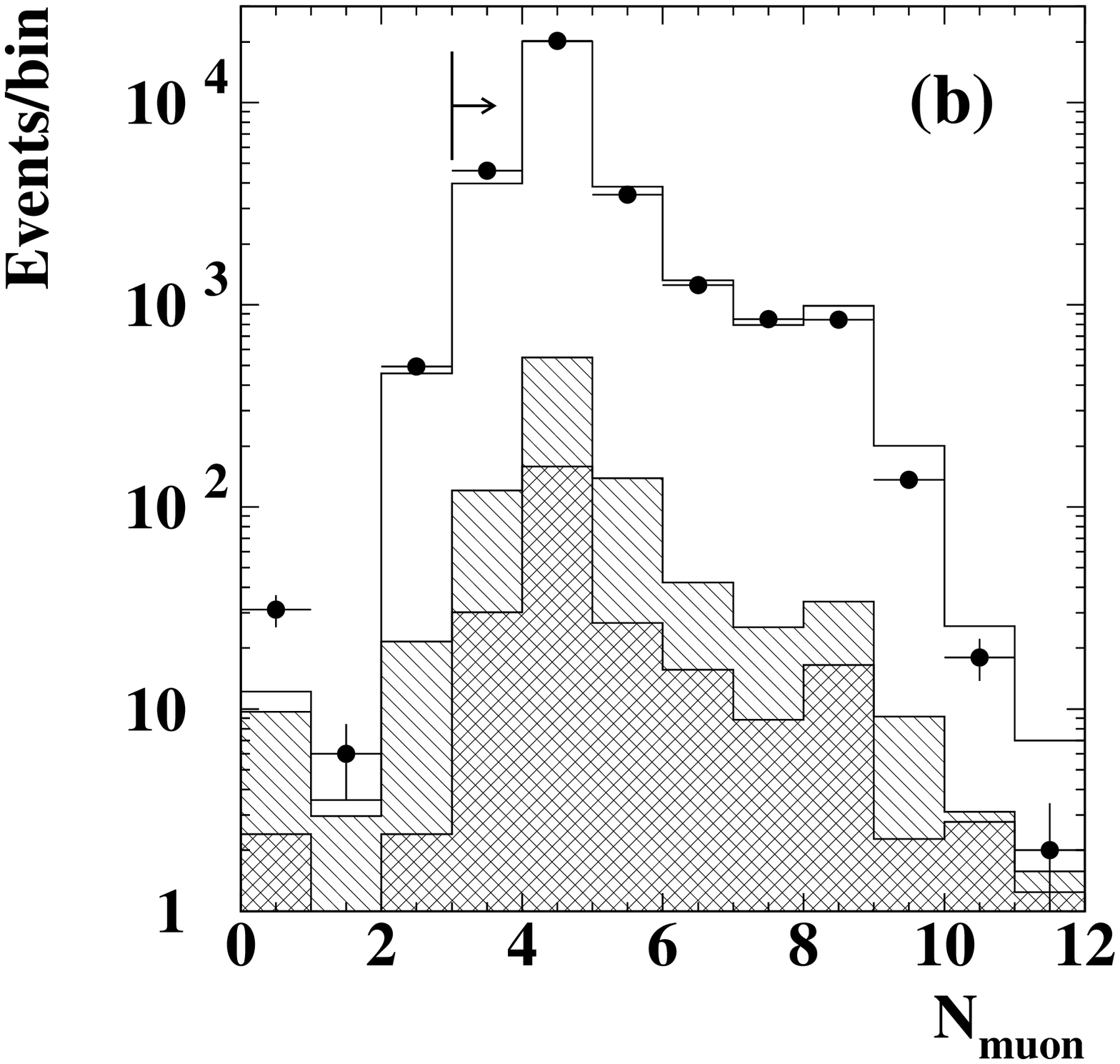,height=8cm}}
\mbox{\epsfig{file=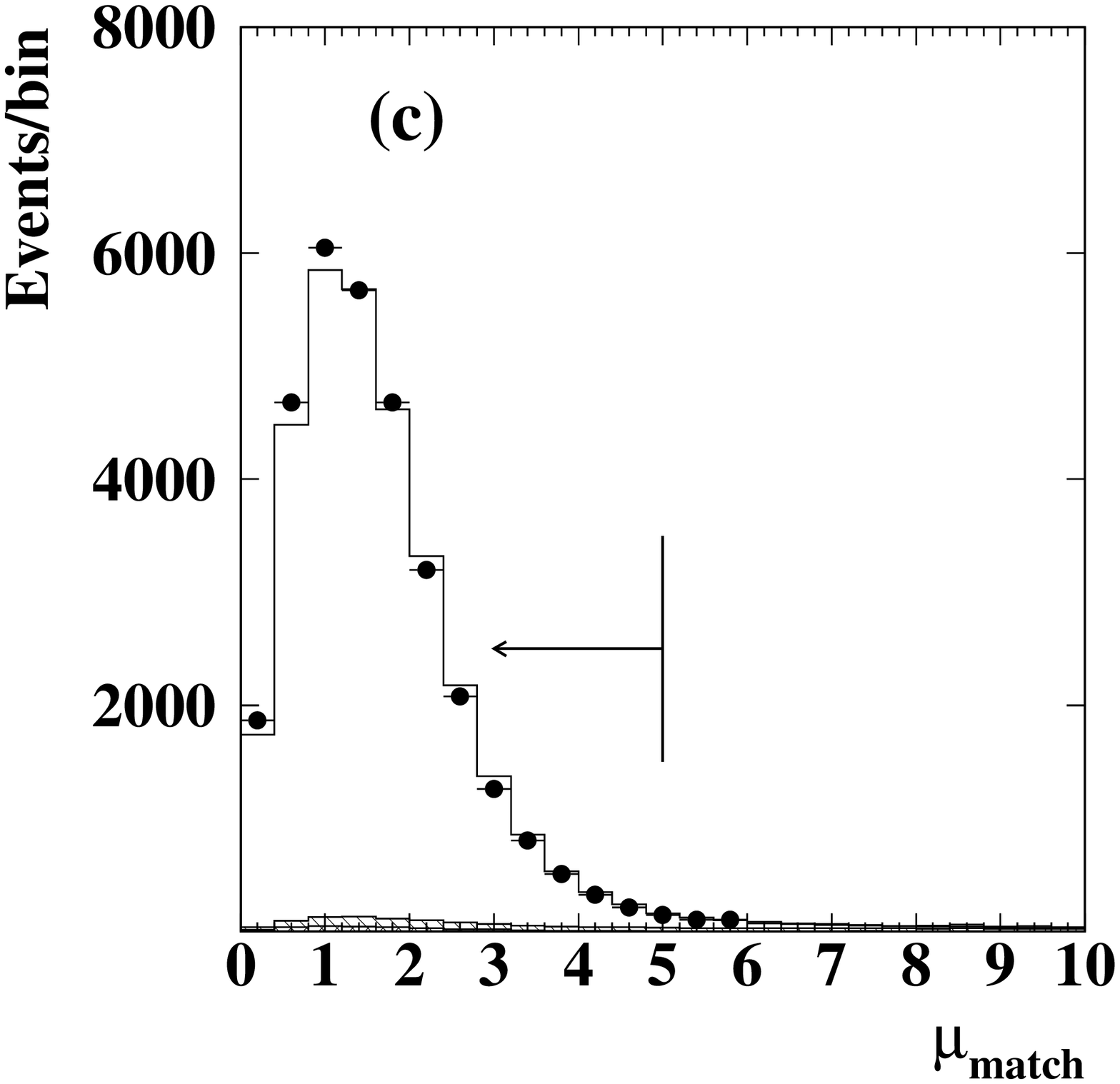,height=8cm}}
\mbox{\epsfig{file=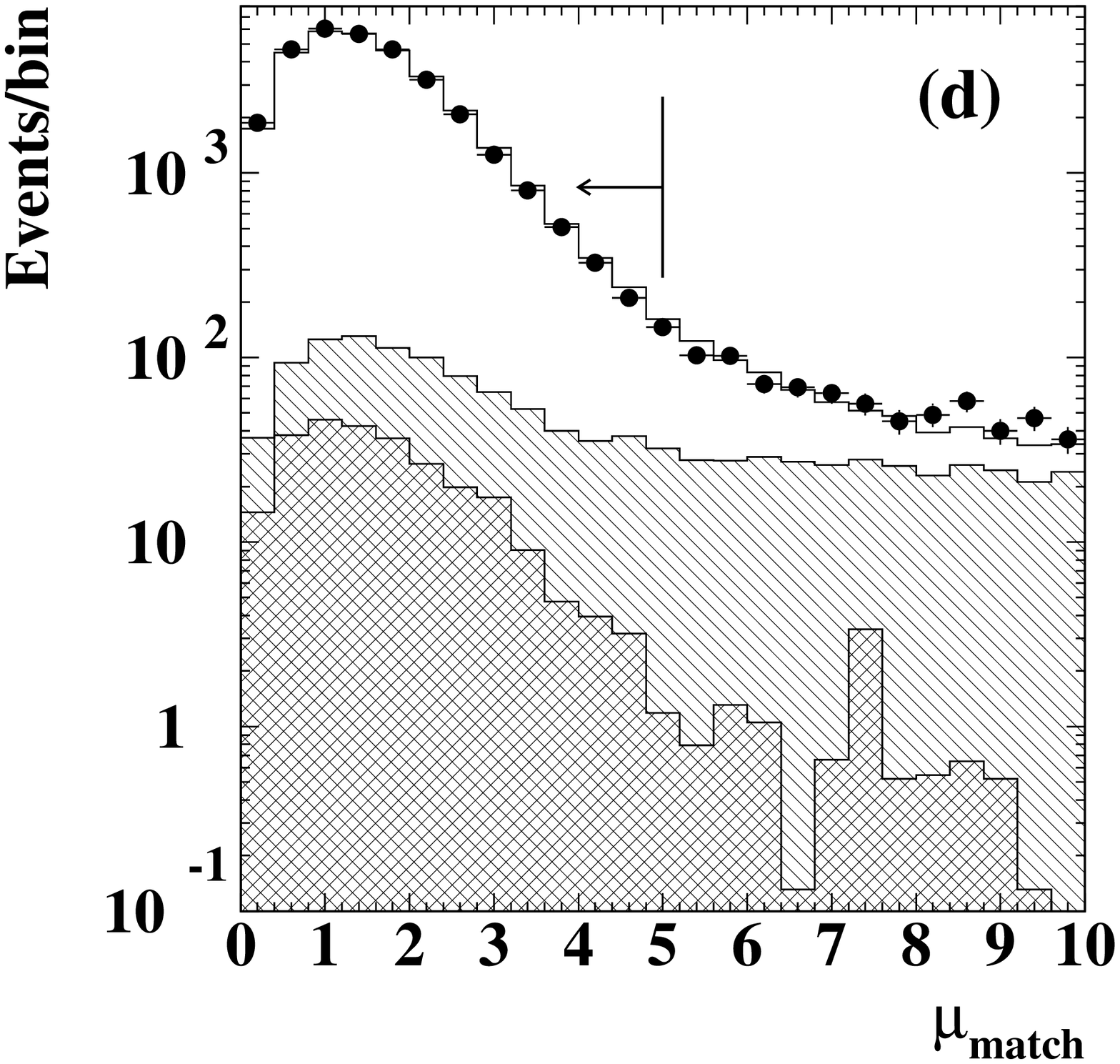,height=8cm}}
\end{center}
\caption[Distributions used in the signal selection.]{\label{cuts} (a,
  b) The
  number of muon layers, $\nmu$, activated by the passage of a charged
particle in the jet, and (c, d) the $\mumatch$
matching between a muon track reconstructed in the tracking chamber
and one reconstructed in the muon chamber.  The jets in each
plot have passed all other selection criteria. The arrows indicate the
accepted regions.  The points are data, the clear histogram
is the Monte Carlo $\taum$ prediction, the singly-hatched histogram is
the Monte Carlo prediction for backgrounds from other $\tau$ decays,
and the cross-hatched histogram is the Monte Carlo prediction for
background from non-$\tau$ sources.}
\end{figure}

Muons are identified by selecting charged particles that produce a signal in
at least three muon chamber layers.  The position of each muon chamber
signal must agree with that of the extrapolated track from the drift
chamber in order for it to be associated with the track.  $\nmu$ is
the number of muon chamber layers activated by a passing particle, and
we require $\nmu > 2$.  Although both the barrel region\footnote{In the muon chambers, the
  barrel region has $|\cos\theta| < 0.68$ and the
  endcaps cover the region where $0.67 < |\cos\theta| < 0.98$.} and
endcap region nominally have four layers of muon 
chambers, there are areas of overlap between different regions which may result 
in more than four layers being activated, as shown in Figure
\ref{cuts} (a) and (b).  The value of the $\nmu$ cut was chosen to minimise the background while retaining signal.
The logarithmic plot shows a small discrepancy between the data and the
Monte Carlo simulation at low values of $\nmu$; however, changing the
value of the cut or removing this criterion entirely does not
significantly affect the branching ratio, as is discussed in Section
\ref{systs}.

  Tracks in the muon
chambers are reconstructed independently from those in the tracking
chamber.  The candidate muon track in the tracking chamber
is typically well-aligned with the corresponding track in the muon
chambers, whereas this is not the case for hadronic $\tau$ decays,
which are the main source of background in the sample.  The majority
of these background jets contain a pion which interacts in the hadronic
calorimeter, resulting in the production of secondary particles which
emerge from the calorimeter and generate signals in the muon chambers, a
process known as pion punchthrough.
Therefore, a
``muon matching'' variable, $\mumatch$, which compares the agreement between
the direction of a track reconstructed in the tracking chamber and
that of the track
reconstructed in the muon chambers, is used to differentiate the signal
$\taum$ decays from
hadronic $\tau$ decays\footnote{$\mumatch$ measures the difference in
  $\phi$ and in $\theta$ between a track reconstructed in the tracking
  chamber and one reconstructed in the muon chambers.  The differences
  are divided by an error estimate and added in quadrature to form a
  $\chisq$-like comparison of the directions.}.
It is required that $\mumatch$ have a value
of less than 5, (see Figure \ref{cuts} (c) and (d)).  The position of
the cut was chosen to minimise the background while retaining signal.

\begin{figure}
\begin{center}
\mbox{\epsfig{file=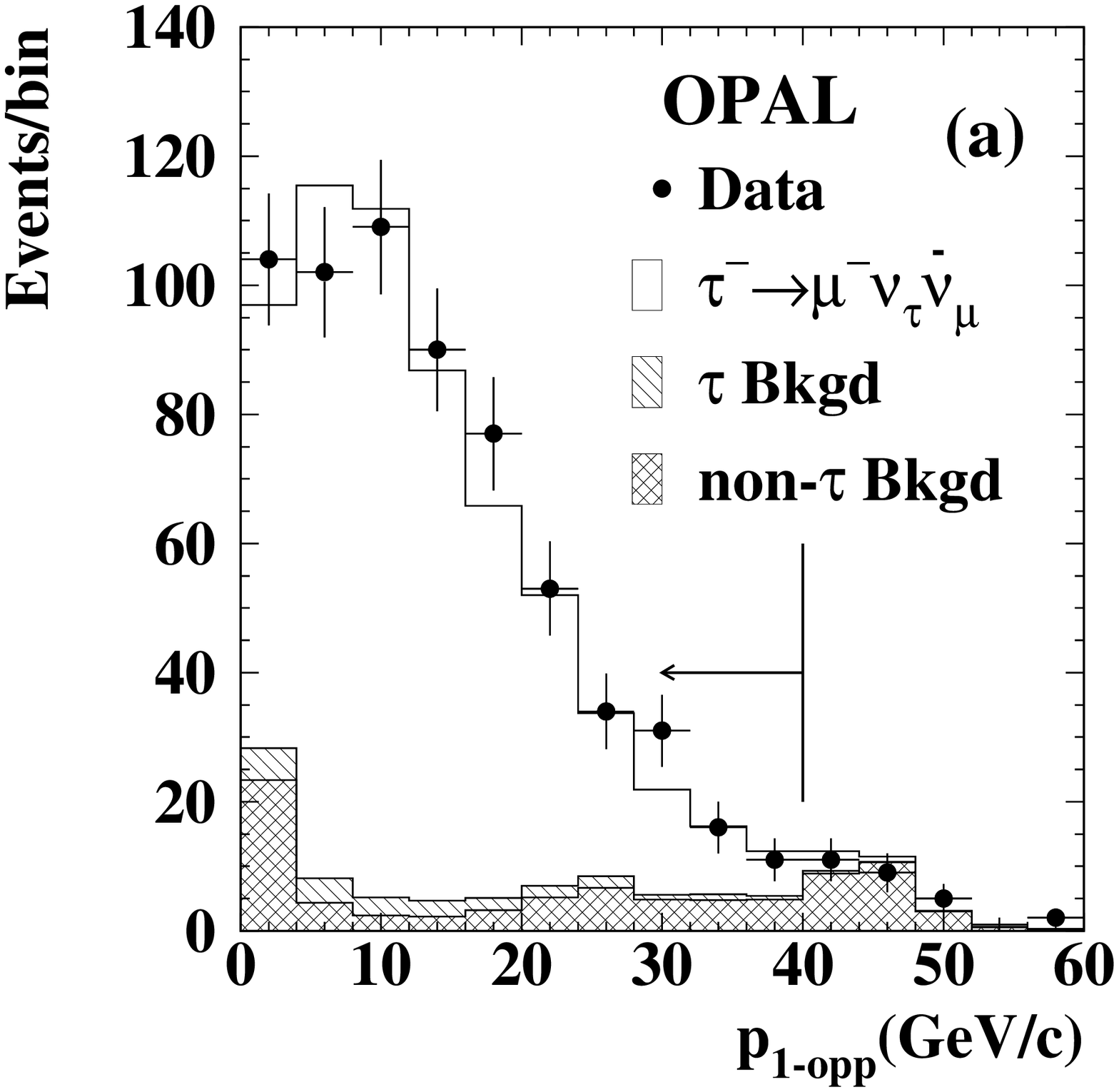,height=8cm}}
\mbox{\epsfig{file=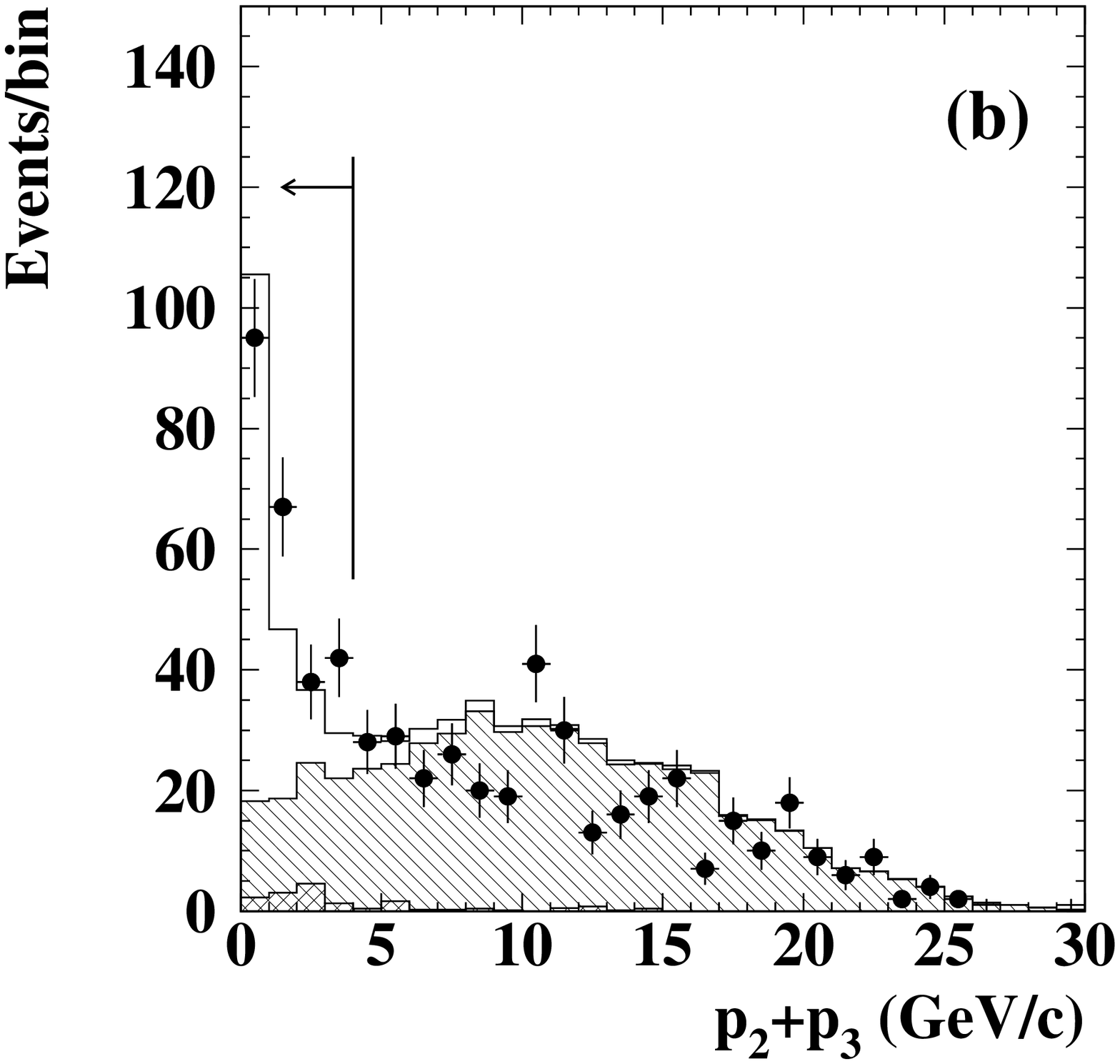,height=8cm}}
\end{center}
\caption[Momentum of 2nd and 3rd tracks.]{\label{ptks} (a) The momentum of
  the highest momentum particle in the opposite jet, $\popp$, where the
candidate muon has a momentum greater than 40 GeV/c, and (b) the combined momentum of the second and
third particles in those jets which have more than one track, for jets
which have passed all
other selection criteria.  The arrows indicate the accepted regions.  The points are data, the clear histogram
is the Monte Carlo $\taum$ prediction, the singly-hatched histogram is
the Monte Carlo prediction for backgrounds from other $\tau$ decays,
and the cross-hatched histogram is the Monte Carlo prediction for
background from non-$\tau$ sources.}
\end{figure}

In order to reduce background from dimuon events, it is required that the
momentum of the highest momentum particle in at least one of the two
jets in the event, i.e. $\pa$ in the candidate jet and $\popp$ in the
opposite jet, must be less than 40 GeV/c (see Figure \ref{ptks} (a)).    

Although the $\taum$ candidates in general are expected to have one
 track, in approximately 2\% of these decays a radiated photon
 converts to an
 $\ee$ pair, resulting in one or two extra tracks in the
tracking chamber.  In order to retain these jets but eliminate
 background jets, it is required that the scalar sum of the 
momenta of the two lower-momentum particles, $\pb$ + $\pc$,  must be
 less than 4 GeV/c (see Figure \ref{ptks} (b)).  In cases where
 there is only one extra track, $\pc$ is taken to be zero.

\begin{figure}
\begin{center}
\mbox{\epsfig{file=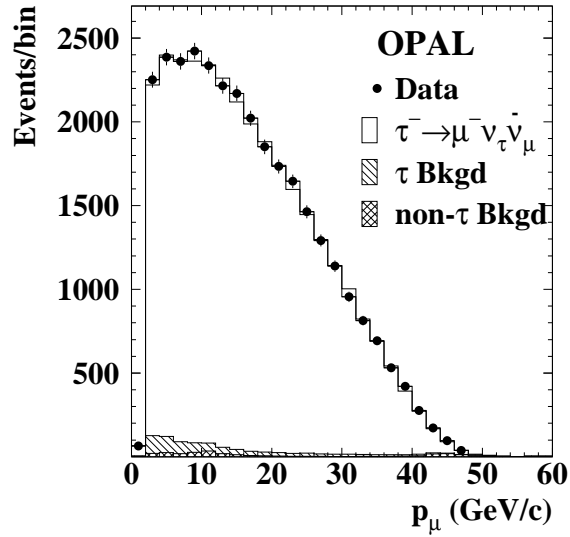,height=8cm}}
\end{center}
\caption[Momentum of the candidate muon.]{\label{ptk} The momentum of
  the candidate muon, $\pmu$, for jets which have passed all
of the selection criteria.  The points are data, the clear histogram
is the Monte Carlo $\taum$ prediction, the singly-hatched histogram is
the Monte Carlo prediction for backgrounds from other $\tau$ decays,
and the cross-hatched histogram is the Monte Carlo prediction for
background from non-$\tau$ sources.}
\end{figure}

The above criteria leave a sample of 31,395 candidate $\taum$ jets.
The quality of the data is illustrated in Figure \ref{ptk}, which
shows the momentum of the candidate muon for jets which satisfy the
$\taum$ selection.
The backgrounds remaining in this sample are discussed in the next section.


\section{\label{taumbk} Backgrounds in the \boldmath $\taum$
\unboldmath sample}
The background contamination in the signal $\taum$ sample stems from
other $\tau$ decay modes and from residual non-$\tau$ background in the
$\tt$ sample.  The procedure used to evaluate the background in the
$\taum$ sample is identical to the one used to evaluate the background
in the $\tt$ sample, which is outlined in Section \ref{tausel}.

\begin{figure}
\begin{center}
\mbox{\epsfig{file=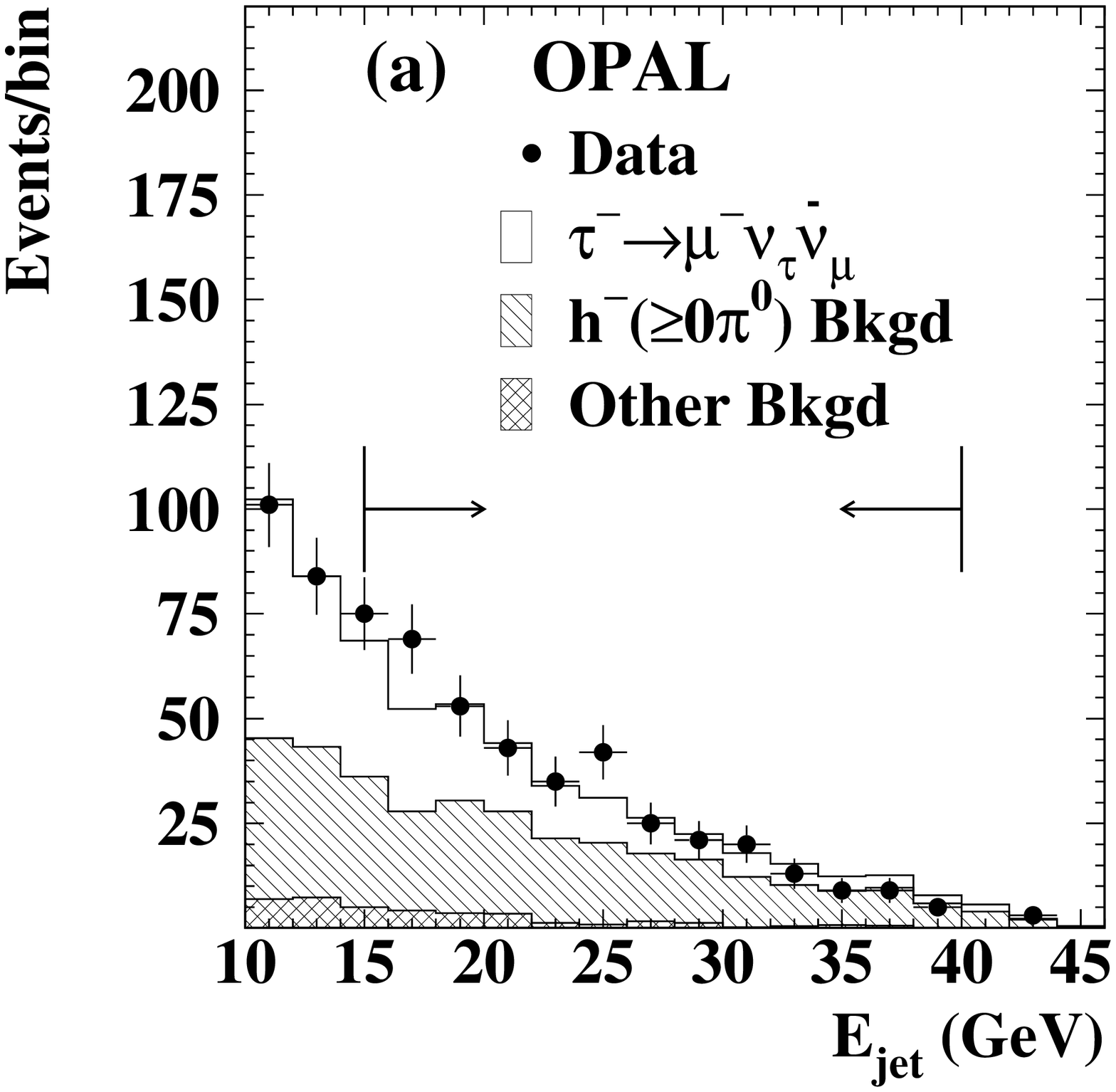,height=8cm}}
\mbox{\epsfig{file=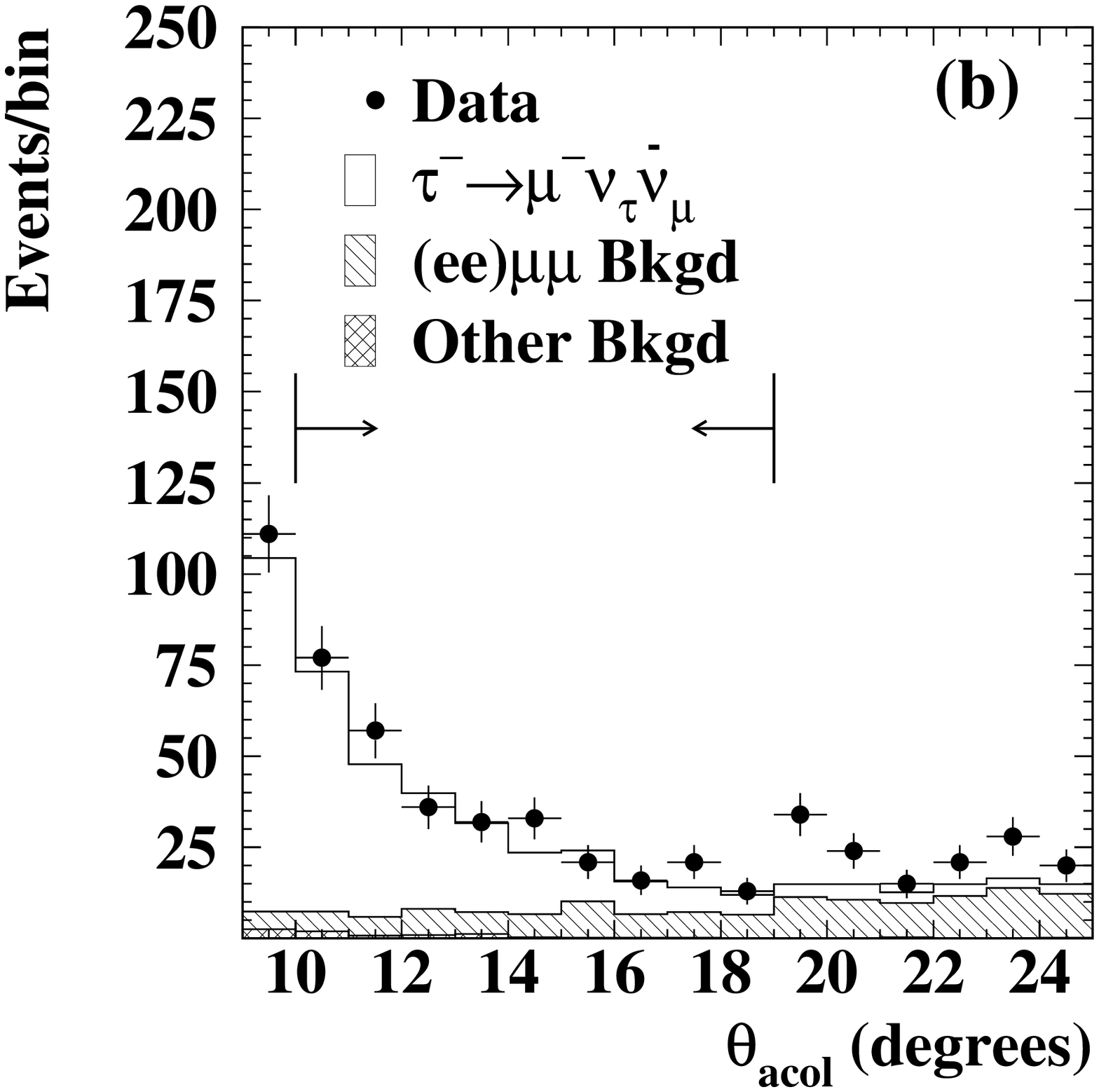,height=8cm}}
\mbox{\epsfig{file=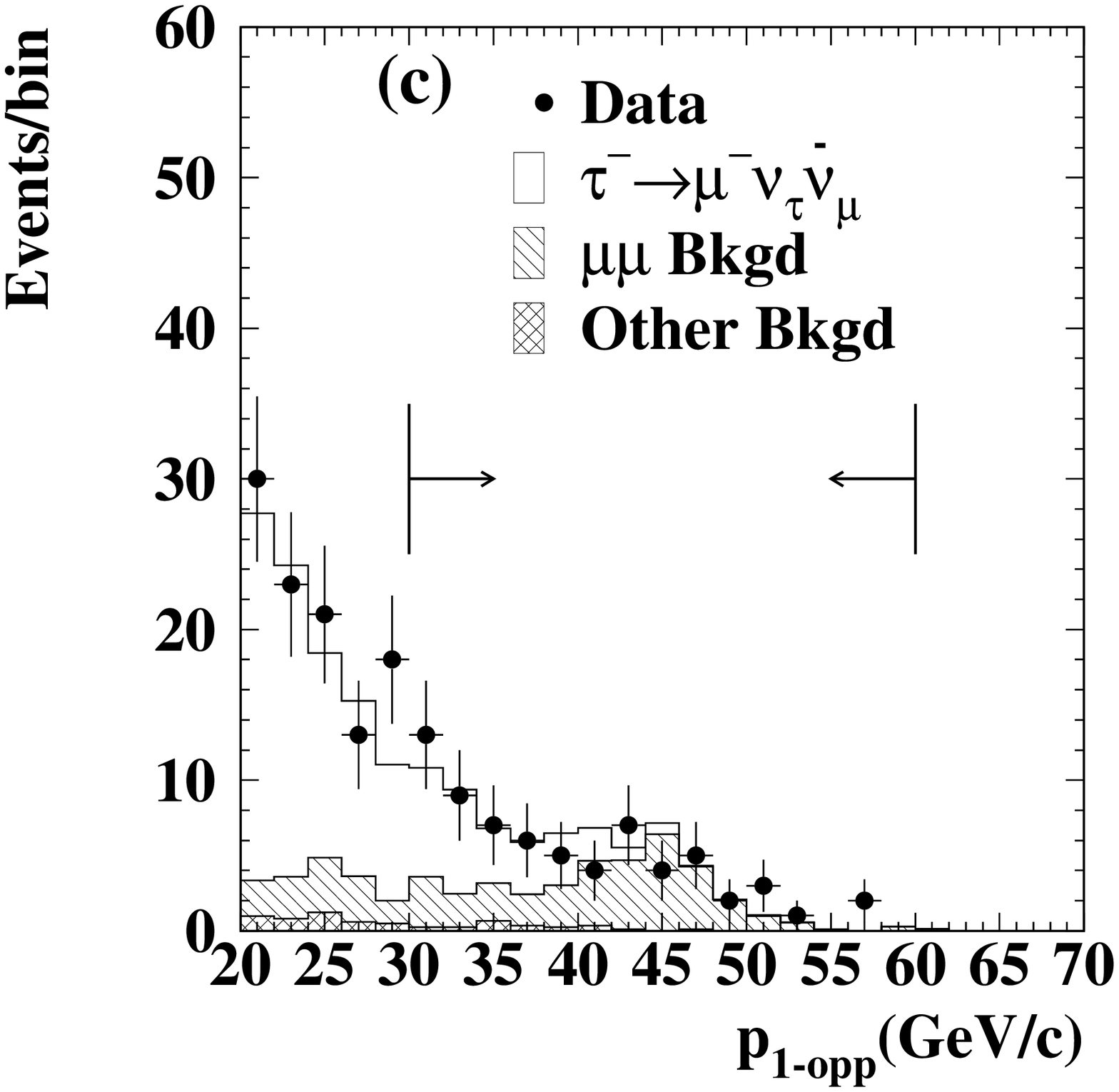,height=8cm}}
\mbox{\epsfig{file=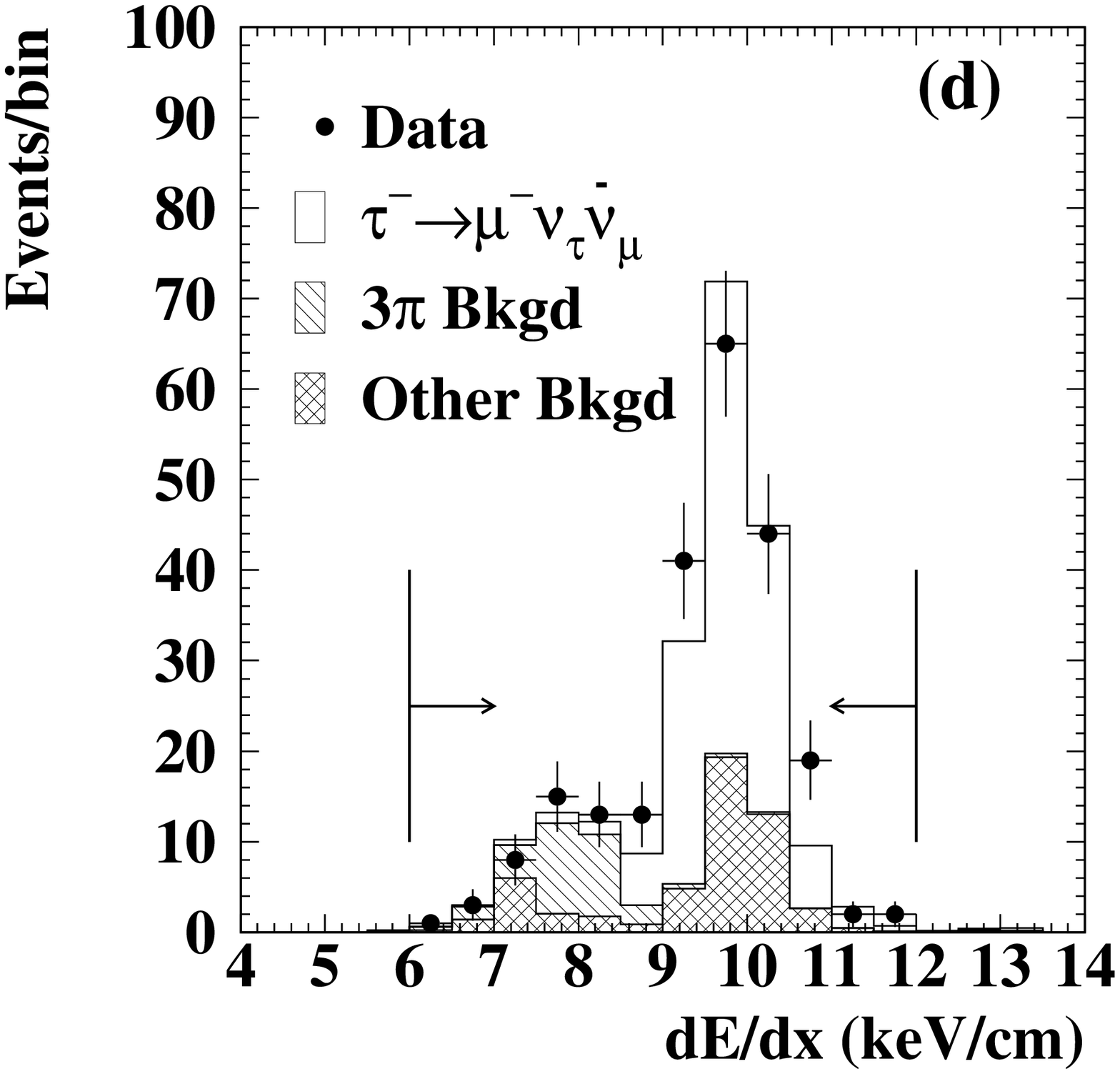,height=8cm}}
\end{center}
\caption[Distributions for background measurements.]{\label{bkdist} The
  distributions used to measure the background in the $\taum$ sample
  are shown after the normalization. The arrows indicate the region that was chosen
to measure each background.  (a) $\ejet$ is the energy measured in the
  electromagnetic calorimeter, (b) $\acol$ is the acollinearity
  angle between the two $\tau$ jets, (c) $\popp$ is the momentum of the
  highest momentum particle in the opposite jet to the $\taum$
  candidate, (d) $\dedxt$ is the rate of energy loss of a particle
  traversing the tracking chamber.  The points are data, the clear histogram
is the Monte Carlo $\taum$ prediction, the singly-hatched histogram is
the Monte Carlo prediction for the type of background being evaluated
  using each distribution,
and the cross-hatched histogram is the Monte Carlo prediction for all
  other types of background.}
\end{figure}

The main backgrounds from other $\tau$ decay modes can be
separated into $\tauhpmay$, and a small number of
$\tauhhh$ jets.  The $\tauhpmay$ decays can pass the $\taum$ selection
when the charged hadron punches through the calorimeters,
leaving a signal in the muon chambers.  The absence or presence of
$\pi^0$s has no impact on whether or not the jet is selected, since
there are over 60 radiation lengths of material in the detector in
front of the muon chambers.  The modelling of this background is
tested by studying $\taum$ jets with large deposits of energy in the
electromagnetic calorimeter. The distribution of jet energy,
$\ejet$,  deposited in the
electromagnetic calorimeter is shown in Figure \ref{bkdist} (a).  The
$\tauhpmay$ fractional background estimate is $0.0225 \pm
0.0016$, of which approximately 75\% includes at least one $\pi^0$.

The main backgrounds resulting from contamination in the $\tt$ sample are
$\reeeemm$ and $\reemm$ events.  The $\reeeemm$ contribution in the $\taum$ sample was
evaluated by fitting the Monte Carlo distribution of the acollinearity
angle, $\acol$,  to that
of the data, where the acollinearity criterion in the $\tt$ selection
which requires that $\acol < 15^{\circ}$ 
has been relaxed, and $\ptot$ is required to be less than 20 GeV/c, as shown in
Figure \ref{bkdist} (b). This
resulted in a fractional background estimate of $0.0052 \pm 0.0026$.
For
this particular background, the quoted uncertainty also takes into account
the spread in the fitted normalization when the range of $\acol$ is
extended to 20 and to 25 degrees. This is motivated by a 
discrepancy between the data and the Monte Carlo simulation 
which can be seen in the region where $\acol > 20^{\circ}$. 

The contribution of dimuon events ($\reemm$) was enhanced in the $\taum$ sample by removing the
requirement that  $\popp < 40$ GeV/c or $\pa < 40$ GeV/c, and instead requiring that
$\pa > 40$ GeV/c. The distribution of $\popp$ was then used to
evaluate the agreement between the data and the Monte Carlo simulation
for this background.  The resulting estimate of the dimuon fractional background in the
$\taum$ sample is $0.0029 \pm 0.0006$.  The corresponding 
distribution is shown in
Figure \ref{bkdist} (c).

Signal events with three tracks are due to radiative $\taum$ decays
where the photon converts in the tracking chamber to an $\ee$ pair, whereas the three-track background
consists mainly of jets with three pions in the final state.
Electrons and pions have different rates of energy loss in the OPAL
tracking chamber, and hence the background can
be isolated from the signal 
by using the rate of energy loss as the particle traverses the tracking
chamber, $\dedxt$, of the second-highest-momentum particle in the
jet.  The Monte Carlo modelling was compared to the data as shown in
Figure \ref{bkdist} (d), yielding a fractional background measurement
of $0.0014 \pm 0.0003$.  

        The remaining background in the $\taum$ sample is almost
        negligible and is estimated from the Monte Carlo simulation.  The total
        estimated fractional background in the $\taum$ sample after the 
        selection is $0.0324 \pm 0.0031$.  The main background
        contributions are summarized in Table
        \ref{tbackgrounds}.

\begin{table} [h]
\begin{center}
\begin{tabular} {|l|l|} \hline
Backgrounds     & Contamination  \\ \hline
$\tauhpmay$     & $0.0225 \pm 0.0016$ \\
$\reeeemm$      & $0.0052 \pm 0.0026$ \\
$\reemm$        & $0.0029 \pm 0.0006$ \\ 
$\tauhhh$       & $0.0014 \pm 0.0003$ \\
Other           & $0.0004 \pm 0.0001$ \\ \hline
Total           & $0.0324 \pm 0.0031$
\\ \hline
\end{tabular}
\caption{\label{tbackgrounds}
The main sources of background in the candidate $\taum$ sample
together with their estimated uncertainties.}
\end{center}
\end{table}

\section{\label{br}The branching ratio}

The $\taum$ branching ratio is given by
\begin{equation}
\label{ebr}
        \mbox{B} =  \frac{N_{(\tau \rightarrow \mu)}}{N_{\tau}} 
             \frac{(1-f_{\mathrm{bk}})}{(1-f_{\tau \mathrm{bk}})}
             \frac{1}{\epsilon_{(\tau \rightarrow \mu)}}
             \frac{1}{F_{\mathrm{b}}},
\end{equation}
where the first term,  $N_{(\tau \rightarrow
  \mu)}/N_{\tau}$, is extracted from the data and is the
number of $\taum$ candidates after the $\taum$ selection, divided by the number of $\tau$ candidates
selected by the $\tt$ selection.  The remaining terms include the estimated fractional backgrounds in the $\taum$ and 
  $\tt$ samples, $f_{\mathrm{bk}}$ and
  $f_{\tau \mathrm{bk}}$, respectively, which must be subtracted off the numerator and
  denominator in the first term of Equation \ref{ebr}.  The evaluation
  of these backgrounds has been discussed in Sections \ref{tausel} and \ref{taumbk}.  The
  efficiency of selecting the $\taum$ jets out of the sample of $\tt$
  candidates is given by $\epsilon_{(\tau \rightarrow \mu)}$.  The Monte Carlo
  prediction of the efficiency is cross-checked using a control
  sample, and will be discussed in Section~\ref{systs}.
  $F_{\mathrm{b}}$ is a bias factor 
which accounts for the fact that the $\tt$ selection does not select
all $\tau$ decay modes with the same efficiency, and will also be
explained in more detail in Section \ref{systs}.   The corresponding
values of these parameters for the $\taum$ selection are
shown in Table \ref{tbrvalues}.  Equation~\ref{ebr} results in a
branching ratio value of 
\[
\mbox{\btaum} = 0.1734 \pm 0.0009 \pm 0.0006,
\] 
where the first error is
statistical and the second is systematic. 
\begin{table} [h]
\begin{center}
\begin{tabular} {|l|l|} \hline
Parameter       & Value     \\ \hline
$N_{(\tau \rightarrow \mu)}$    & 31,395 \\
$N_{\tau}$      & 193,796         \\
$f_{\mathrm{bk}}$        & $0.0324 \pm 0.0031$    \\
$f_{\tau \mathrm{bk}}$   & $0.0106 \pm 0.0007$  \\
$\epsilon_{(\tau \rightarrow \mu)}$     & $0.8836 \pm 0.0021$ \\ 
$F_{\mathrm{b}}$         & $1.0339 \pm 0.0020$     \\ \hline
B($\taum$)      & $0.1734 \pm 0.0009(stat) \pm 0.0006(syst)$ \\ \hline
\end{tabular}
\caption{\label{tbrvalues}
Values of the quantities used in the calculation of \btaum.}
\end{center}
\end{table}

\subsection{\label{systs} Systematic checks}

The statistical uncertainty in the branching ratio is taken to be the binomial error in the
uncorrected branching ratio, $N_{(\tau \rightarrow \mu)}/N_{\tau}$.  The systematic errors
include the contributions associated with the Monte Carlo modelling of
each of the main sources of background in the $\taum$ sample, the error
in the efficiency, the error in the
background in the $\tt$ sample, and the error in the bias
factor.  These
errors are listed in Table \ref{tbrvalues} and their contribution to
the error in the branching ratio is shown in Table \ref{tbrerrors}.
The errors in the backgrounds have already been discussed in Sections
\ref{tausel} and \ref{taumbk}.  A discussion of the error in the
efficiency and in the bias factor follows.

A second sample of $\taum$ data candidates was selected using
information from the tracking chamber plus the electromagnetic and
hadronic calorimeters.  The selection looks for jets with
one to three tracks satisfying $\pb + \pc < 4$ GeV/c, and which leave
little energy in the electromagnetic or hadronic calorimeters but
still leave an observable signal in several layers of the hadronic calorimeter.  This yields a sample of 28,042
$\taum$ jets and results in a
branching ratio of 0.1730 with a measured
fractional background of 0.0396 and an efficiency of 0.7853.  The candidates selected using this
{\it calorimeter} selection are highly correlated with those selected
for the main branching ratio analysis using the {\it tracking} selection, even though the two selection
procedures are largely independent.  Because of the high level of
correlation, the advantage of combining the two selection methods is
negligible; however, the calorimeter selection is very useful for
producing a
control sample of $\taum$ jets which can be used for systematic checks.  

A potentially important source of systematic error in the analysis is
the Monte Carlo modelling of the selection efficiency.  In order to
estimate the  error on the efficiency, both
data and Monte Carlo simulated jets are required to satisfy the
calorimeter selection criteria.  This produces two control samples of candidate $\taum$
jets,  one which is data, and one which is Monte Carlo simulation.
  The efficiency of the tracking selection is
then evaluated as the fraction of jets in the calorimeter sample which pass
the tracking selection.  The ratio of the efficiency found using the
data to the efficiency found using the Monte Carlo simulation is
$1.0002 \pm 0.0024$.  The uncertainty in the ratio was taken as the systematic error in the
$\taum$ selection efficiency.

Further checks of the Monte Carlo modelling are made by varying each of the selection
criteria and noting the resulting changes in
the branching ratio.  The requirement on the number of tracks was changed to allow only
one track in the jet, in order to remove the radiative decays with
photon conversions.  This
was found to change the branching ratio by 0.0003.  
Changing the
requirement on $\nmu$ from two to one resulted in a branching ratio
change of 0.0002.  Removing this criterion entirely resulted in a
change of 0.0003.  Varying the $\mumatch$ value of the match between a
tracking chamber track and a muon chamber track by $\pm 1/2$ resulted
in changes of 0.0002.  The requirement on $\popp$ was changed by $\pm
2$ GeV/c and resulted in a change of 0.00001.  Removing the requirement
of $\popp$ entirely results in a similar change.  All of these
changes are within
the systematic uncertainty that has already been assigned due to the background and
efficiency errors, which are equivalent to an uncertainty in the
branching ratio of 0.0005.  Thus one has confidence that the error in
the modelling of the background and the signal does not exceed the
error already quoted. 

The $\tau$ Monte Carlo simulations create events for the different
$\tau$ decay modes in accordance with the measured $\tau$ decay branching ratios \cite{pdg}.  However, the $\tt$
selection does not select each $\tau$ decay mode with equal
efficiency.  This can introduce a bias in the measured value of
B($\taum$).  The $\tt$ selection bias factor, $F_{\mathrm{b}}$, measures the
degree to which the $\tt$ selection favours or suppresses the decay
$\taum$ relative to other $\tau$ decay modes.  It is defined as the
ratio of the fraction of $\taum$ decays in a sample of $\tau$ decays
after the $\tt$ selection is applied, to the fraction of $\taum$
decays before the selection.  The dependence of the bias factor on
\btaum \, was checked by varying the branching ratio within the
uncertainty of 0.0007 given in reference \cite{pdg}.  This resulted in negligible changes
in the bias factor.  In addition, extensive studies of
systematic errors in the bias factor have been made in previous OPAL $\tau$-decay
analyses \cite{steve, john}, including rescaling the
centre-of-mass energy and then recalculating the bias factor, and 
smearing some Monte Carlo variables  and then again recalculating the
bias factor.  These checks have indicated that the
systematic effects do not contribute to the uncertainty in a
significant manner compared with the statistical uncertainty, and so
we have not included a systematic component in the error. 

\begin{table} [h]
\begin{center}
\begin{tabular} {|l|c|} \hline
Source       & Absolute error     \\ \hline
$\epsilon_{(\tau \rightarrow \mu)}$  & 0.00040 \\
$F_{\mathrm{b}}$ & 0.00034         \\
$f_{\mathrm{bk}}$        & 0.00030    \\
$f_{\tau \mathrm{bk}}$   & 0.00012  \\ \hline
Total      & 0.00062 \\ \hline
\end{tabular}
\caption{\label{tbrerrors}
Contributions to the total branching ratio absolute systematic
uncertainty.  The uncertainty in $f_{\mathrm{bk}}$ has been adjusted
to take into account correlations between the backgrounds in the $\tt$
and $\taum$ samples.}
\end{center}
\end{table}

\section{Discussion}

The value of B$(\taum)$ obtained in this analysis can be used in
conjunction with the previously measured OPAL value of B$(\taue)$ to
test various aspects of the Standard Model.  For example, the Standard
Model assumption of lepton universality implies that the
coupling of the W boson to all three generations of
leptons is identical.  The leptonic $\tau$ decays have already provided some of the most
stringent tests of this hypothesis (see, for example, \cite{steve}).
With the improved precision of
B$(\taum)$ presented in this paper, it is worth testing this
assumption again. In addition, the leptonic $\tau$ branching ratios can be used to
measure the Michel parameter $\eta$, which can be used 
to set a limit on the mass of the charged Higgs particle in the
Minimal Supersymmetric Standard Model.  These
topics are discussed below.

\subsection{Lepton universality}

The Standard Model assumption of
lepton universality implies that the coupling constants $\ge$, $\gm$,
and $\gt$  are identical, thus the ratio $\gm/\ge$ is expected to be
unity.  This can be tested experimentally by taking the ratio of the
corresponding branching ratios, which yields
\begin{equation}
\label{egmge}
        \frac{\mbox{\btaum}}{\mbox{B}(\taue)}
                = \frac{\gm^2}{\ge^2} \left[\frac{f\left(
                \frac{m^2_{\mu}}{m^2_{\tau}}\right)}{f\left(
                \frac{m^2_{\mathrm{e}}}{m^2_{\tau}}\right)}\right]
\end{equation}
where $f(m^2_{\mathrm{e}}/m^2_{\tau}) =  1.0000$ and
$f(m^2_{\mu}/m^2_{\tau}) = 0.9726$ are the corrections for the masses
of the final state leptons \cite{marciano}.  We use Equation \ref{egmge} to
compute the coupling constant ratio, which, with the value
of B($\taum$) from this work and
the OPAL measurement of B($\taue$) = $0.1781 \pm 0.0010$ \cite{steve},
yields
\[
        \frac{\gm}{\ge} = 1.0005 \pm 0.0044,
\]
in good agreement with expectation.  The OPAL measurements of the branching ratios
B($\taue$) and B($\taum$) are assumed to be uncorrelated.

\begin{figure}
\begin{center}
\mbox{\epsfig{file=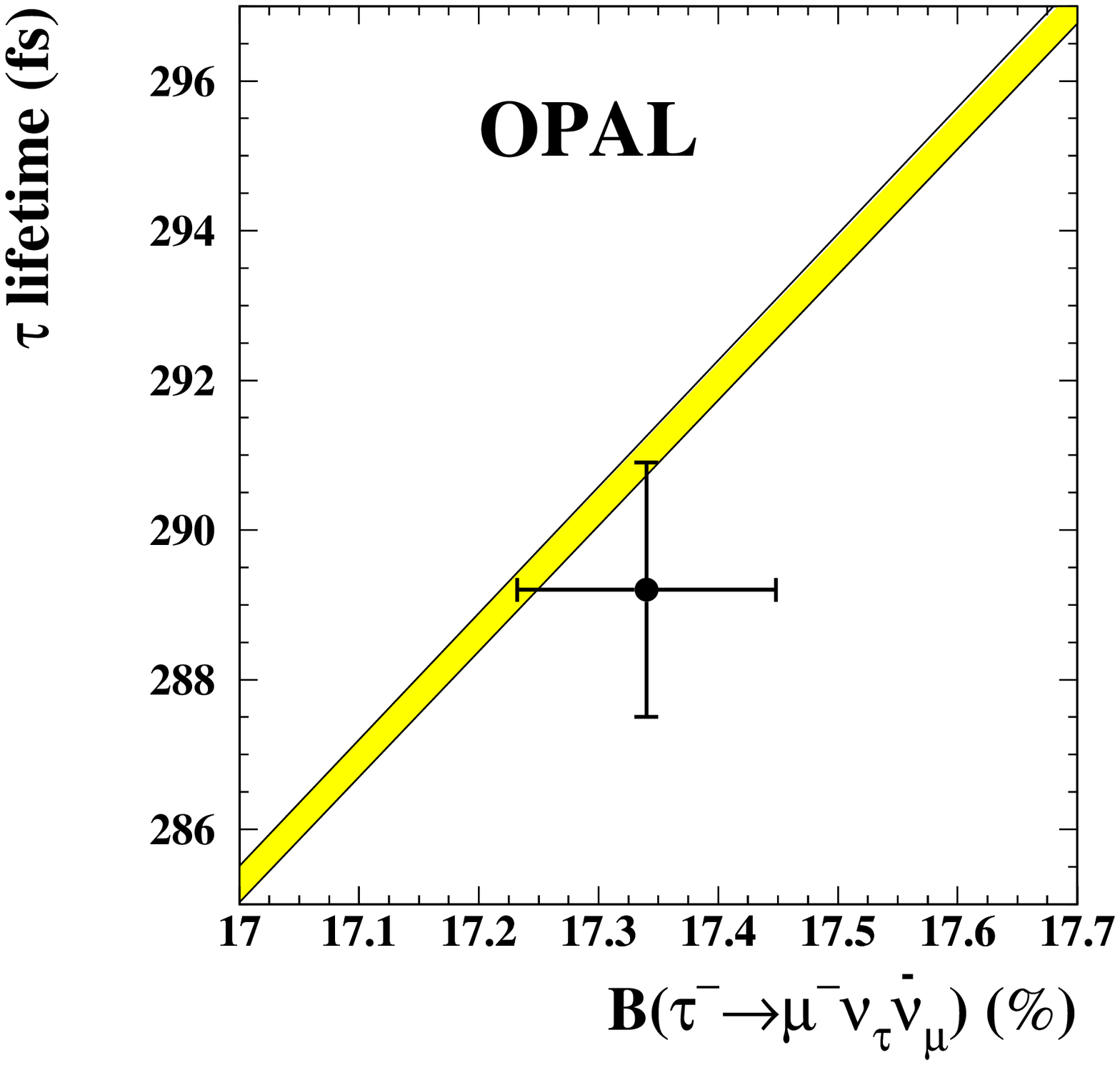,height=16cm}}
\end{center}
\caption[The $\tau$ lifetime vs B($\taum$).]{\label{ttauvsbr} The lifetime of the $\tau$ vs the $\taum$ branching ratio.  The
band is the Standard Model expectation, and its width is determined by
the uncertainty in the mass of the $\tau$ \cite{bes}.  The point with
error bars is the OPAL measurement of the $\tau$ lifetime \cite{ttau} and the branching ratio determined in this work.}
\end{figure}

In addition, the $\taum$ branching ratio can be used in conjunction
with the muon and $\tau$ masses and lifetimes to test lepton
universality between the first and third lepton generations, yielding
the expression
\begin{equation}
\label{egtge}
        \frac{\gt^2}{\ge^2} = \mbox{\btaum}
        \frac{m_{\mu}^5}{m_{\tau}^5}
        \frac{\tau_{\mu}}{\tau_{\tau}}
        \, \, \frac{f\left(\frac{m^2_{\mathrm{e}}}{m^2_{\mu}}\right)}
        {f\left(\frac{m^2_{\mu}}{m^2_{\tau}}\right)} \, \,
        \frac{(1+ \delta_{\mathrm{RC}}^{\mu})}{(1+ \delta_{\mathrm{RC}}^{\tau})} \, \, \, .
\end{equation}
The values $(1+\delta_{\mathrm{RC}}^{\tau})=0.99597$ and
$(1+\delta_{\mathrm{RC}}^{\mu})=0.99576$, which take into account
photon radiative corrections and leading order W propagator
corrections, and $f(m^2_{\mathrm{e}}/m^2_{\mu}) =  0.9998$,
are obtained from reference \cite{marciano}.  Using the OPAL value for
the $\tau$ lifetime, $\tau_{\tau} =
289.2 \pm 1.7 \pm 1.2$ fs \cite{ttau}, the BES collaboration value for
the $\tau$ mass, $m_{\tau} = 1777.0 \pm 0.3$ MeV/$c^2$ \cite{bes}, and the Particle
Data Group \cite{pdg} values for the muon mass, $m_{\mu}$, and muon lifetime,
$\tau_{\mu}$, we obtain
\[
        \frac{\gt}{\ge} = 1.0031 \pm 0.0048,
\]      
again in good agreement with the Standard
Model assumption of lepton universality.  If one assumes lepton
universality, then Equation \ref{egtge} can be rearranged to express the $\tau$ lifetime as a function of the
branching ratio \btaum.
The resulting relationship is plotted in Figure \ref{ttauvsbr}.

\subsection{Michel parameter $\eta$ and the charged Higgs mass}

The leptonic $\tau$ branching ratios can be used to probe the Lorentz
structure of the matrix element through the Michel parameters
\cite{rainer,michel},
$\eta$, $\rho$, $\xi$, and $\delta$, which parameterize the shape of
the $\tau$ leptonic decay spectrum. In the Standard Model V-A
framework, $\eta$ takes the value zero.  A non-zero value of $\eta$
would contribute an extra term to the leptonic $\tau$ decay widths.
This effect potentially would be measurable by taking the ratio of
branching ratios, as in Equation \ref{efindeta} \cite{stahl},
\begin{equation}
\label{efindeta}
\frac{\mbox{\btaum}}{\mbox{B}(\taue)} = 0.9726 \left(1 + 4 \eta \frac{m_{\mu}}{m_{\tau}}\right).
\end{equation}
The \btaum \, result presented here, together with 
the OPAL measurement of B$(\taue)$ \cite{steve} and Equation
\ref{efindeta}, then results in a value of $\eta = 0.004 \pm 0.037$.
This can be compared with a previous OPAL result of $\eta = 0.027 \pm
0.055$ \cite{rainer} which has been obtained by fitting the $\tau$ decay spectrum. 

In addition, a non-zero $\eta$ may imply the presence of scalar
couplings, such as those predicted in the Minimal Supersymmetric
Standard Model. The dependence of $\eta$ upon the mass of the charged
Higgs particle in this model, $m_{\mathrm{H}^{\pm}}$, can be
approximately written as \cite{stahl,dova}
\begin{equation}
\label{emasshiggs}
\eta = -\frac{m_{\tau}m_{\mu}}{2}
\left(\frac{\tan\beta}{m_{\mathrm{H}^{\pm}}}\right)^2,
\end{equation}
where $\tan\beta$ is the ratio of the vacuum expectation values of the two
Higgs fields.  Thus, $\eta$ can be used to place constraints on the mass of the
charged Higgs.  A one-sided 95\% confidence limit using the
$\eta$ evaluated in this work gives a value of $\eta > -0.057$, and
a model-dependent limit on
the charged Higgs mass of $m_{\mathrm{H}^{\pm}} > 1.28 \tan\beta$.

\section{Conclusions}

OPAL data collected at energies near the Z$^0$ peak have been used
to determine the  $\taum$ branching ratio, which is found to be
\[
\mbox{\btaum} = 0.1734 \pm 0.0009(stat) \pm 0.0006(syst).
\]
This is the most precise measurement to date, and is consistent with the previous OPAL measurement
\cite{opalt2m} and with previous results from other experiments \cite{pdg}.

The branching ratio measured in this analysis, in conjunction with the OPAL
$\taue$ branching ratio measurement, has been used to verify lepton
universality at the level of 0.5\%.  Although lepton universality has
been tested to precisions of 0.2\% using pion decays, the
scalar nature of pions constrains the mediating W boson to be
longitudinal,  whereas $\tau$ decays involve transverse W bosons, making
these two universality tests
potentially sensitive to different types of new physics.

In addition, these branching
ratios have been used to obtain a value for the Michel parameter
$\eta = 0.004 \pm 0.037$, which in turn has been used to place a limit on the mass of the charged Higgs boson, $m_{\mathrm{H}^{\pm}} > 1.28
\tan\beta$, in the Minimal Supersymmetric Standard Model.  
This result is complementary to that from another recent OPAL analysis
\cite{btotau}, where a limit of $m_{\mathrm{H}^{\pm}} > 1.89
\tan\beta$ has been obtained from the decay $b \rightarrow \tau^-
\bar{\nu}_{\tau} \mathrm{X}$. 

\appendix
\par
\section*{Acknowledgements}
\par
We particularly wish to thank the SL Division for the efficient operation
of the LEP accelerator at all energies
 and for their close cooperation with
our experimental group.  In addition to the support staff at our own
institutions we are pleased to acknowledge the  \\
Department of Energy, USA, \\
National Science Foundation, USA, \\
Particle Physics and Astronomy Research Council, UK, \\
Natural Sciences and Engineering Research Council, Canada, \\
Israel Science Foundation, administered by the Israel
Academy of Science and Humanities, \\
Benoziyo Center for High Energy Physics,\\
Japanese Ministry of Education, Culture, Sports, Science and
Technology (MEXT) and a grant under the MEXT International
Science Research Program,\\
Japanese Society for the Promotion of Science (JSPS),\\
German Israeli Bi-national Science Foundation (GIF), \\
Bundesministerium f\"ur Bildung und Forschung, Germany, \\
National Research Council of Canada, \\
Hungarian Foundation for Scientific Research, OTKA T-029328, 
and T-038240,\\
The NWO/NATO Fund for Scientific Reasearch, the Netherlands.\\

\newpage


\newpage
\begin{thebibliography} {99}

\bibitem{steve} OPAL Collaboration, G. Abbiendi \etal, 
Phys. Lett. {\bf B447} (1999) 134.

\bibitem{opalt2m} OPAL Collaboration, R. Akers \etal, Z. Phys. {\bf
C66} (1995) 543. 

\bibitem{rainer} OPAL Collaboration, K. Ackerstaff \etal, 
Eur. Phys. J. {\bf C8} (1999) 3.

\bibitem{radt2m} OPAL Collaboration, G. Alexander \etal, Phys. Lett. {\bf B388} (1996) 437.

\bibitem{opal} OPAL Collaboration, K. Ahmet \etal, Nucl. Inst. and
Meth. {\bf A305} (1991) 275; \\
OPAL Collaboration, P.P. Allport \etal, Nucl. Inst. and Meth. {\bf A324} (1993) 34; \\
OPAL Collaboration, P.P. Allport \etal, Nucl. Inst. and Meth. {\bf A346} (1994) 476.

\bibitem{koralz} S. Jadach, B.F.L. Ward, and Z. W\c{a}s,
  Comp. Phys. Comm. {\bf 79} (1994) 503.

\bibitem{tauola} S. Jadach \etal, Comp. Phys. Comm. {\bf 76} (1993)
361.

\bibitem{opalsim} J. Allison \etal, Nucl. Inst. and Meth. {\bf A317}
(1992) 47.

\bibitem{jetset} T. Sj\"{o}strand, Comp. Phys. Comm. {\bf 82} (1994)
74.

\bibitem{bhwide} S. Jadach, W. Placzek, and B.F.L. Ward, 
Phys. Lett. {\bf B390} (1997) 298.

\bibitem{vermaseren} R. Bhattacharya, J. Smith, and G. Grammer, 
Phys. Rev. {\bf D15} (1977) 3267; \\
 J.A.M. Vermaseren, and G. Grammer, Phys. Rev. {\bf D15} (1977) 3280.

\bibitem{tausel} OPAL Collaboration, G. Alexander \etal, 
Phys. Lett. {\bf B266} (1991) 201; \\
 OPAL Collaboration, P. Acton \etal, Phys. Lett. {\bf B288} (1992) 373.

\bibitem{lineshape} OPAL Collaboration, G. Abbiendi \etal, Eur. Phys. J. {\bf C19} (2001) 587.

\bibitem{jetalgo} OPAL Collaboration, G. Alexander \etal, Z. Phys. {\bf C52} (1991) 175.



\bibitem{pdg} Particle Data Group, D.E. Groom \etal, Eur.
Phys. J. {\bf C15} (2000) 1.

\bibitem{john} OPAL Collaboration, K. Ackerstaff \etal, 
  Eur. Phys. J. {\bf C4} (1998) 193.



\bibitem{marciano}W.J. Marciano and A. Sirlin, Phys. Rev. Lett. {\bf
61} (1988) 1815.

\bibitem{ttau}OPAL Collaboration, G. Alexander \etal, 
Phys. Lett. {\bf B374} (1996) 341.

\bibitem{bes}BES Collaboration, J.Z. Bai \etal, Phys. Rev.
{\bf D53} (1996) 20. 


\bibitem{michel} L. Michel, Proc. Phys. Soc. {\bf A63} (1950) 514.

\bibitem{stahl} A. Stahl, Phys. Lett. {\bf B324} (1994) 121 and
references therein.

\bibitem{dova}M.T. Dova, J. Swain and L. Taylor, 
Nucl. Phys. B (Proc. Suppl.) {\bf 76} (1999) 133.

\bibitem{btotau}OPAL Collaboration, G. Abbiendi \etal,
  Phys. Lett. {\bf B520} (2001) 1.


\end{thebibliography}
\end{document}